\documentclass[showpacs,preprintnumbers,amsmath,amssymb]{revtex4}
\usepackage{graphicx}
\usepackage{dcolumn}
\usepackage{bm}
\usepackage{epsf}
\newcommand{\p}{\partial}

\newcommand{\occ}{\overline{c}}
\newcommand{\al}{\alpha}

\newcommand{\MSbar}{\overline{\mathrm{MS}}}

\begin{document}
\preprint{LTH-733}
\title{UV finiteness of $3D$ Yang-Mills theories with a regulating mass in the Landau gauge}
\author{D. Dudal$^a$}
 \altaffiliation{D.~Dudal is a postdoctoral fellow of the
\emph{Special Research Fund} of Ghent University}
\email{david.dudal@ugent.be}
\author{J.A. Gracey$^{b}$}
    \email{jag@amtp.liv.ac.uk}
\author{R.F. Sobreiro$^c$}
 \email{sobreiro@dft.if.uerj.br}
\author{S.P. Sorella$^c$}
\altaffiliation{Work supported by FAPERJ, Funda{\c c}{\~a}o de
Amparo {\`a} Pesquisa do Estado do Rio de Janeiro, under the program
{\it Cientista do Nosso Estado}, E-26/151.947/2004.}
\email{sorella@uerj.br}
\author{H. Verschelde$^a$}
 \email{henri.verschelde@ugent.be}
 \affiliation{\vskip 0.1cm $^a$ Ghent University
\\ Department of Mathematical
Physics and Astronomy \\ Krijgslaan 281-S9 \\ B-9000 Gent,
Belgium\\\\
\vskip 0.1cm $^b$ Theoretical Physics Division\\ Department of Mathematical Sciences\\ University of Liverpool\\ P.O. Box 147, Liverpool, L69 3BX, United Kingdom\\\\
\vskip 0.1cm $^c$ UERJ - Universidade do Estado do Rio de
Janeiro\\Rua S\~{a}o Francisco Xavier 524, 20550-013
Maracan\~{a}\\Rio de Janeiro, Brasil}
\begin{abstract}
We prove that three-dimensional Yang-Mills theories in the Landau
gauge supplemented with a infrared regulating, parity preserving
mass term are ultraviolet finite to all orders. We also extend this
result to the Curci-Ferrari gauge.
\end{abstract}
\maketitle

\section{Introduction}In a previous paper \cite{Dudal:2004ch}, we started to investigate
three-dimensional Yang-Mills theories in the Landau gauge,
supplemented with a parity preserving mass term. We recall that the
$SU(N)$ Yang-Mills action in $d$-dimensional Euclidean space time is
given by
\begin{equation}\label{eq1}
    S=S_\mathrm{YM}+S_\mathrm{GF+FP}\;,
\end{equation}
where the classical action reads
\begin{eqnarray}\label{eq2}
    S_{\mathrm{YM}}&=&\int
    d^dx\left(\frac{1}{4}F_{\mu\nu}^aF_{\mu\nu}^a\right)\;,\nonumber\\
F_{\mu\nu}^a&=&\p_\mu A_\nu^a-\p_\mu
A_\nu^a-gf^{abc}A_\mu^bA_\nu^c\;,
\end{eqnarray}
and the gauge fixing part
\begin{equation}\label{eq3}
    S_{\mathrm{GF+FP}}=\int d^dx\left(b^a\p_\mu A^{\mu a}+c^a\p_\mu
    D_\mu^{ab}\occ^b\right)\;,
\end{equation}
enforces the Landau gauge $\p_\mu A_{\mu}^a=0$. $b^a$ is the
Lagrange multiplier and $c^a$, $\occ^a$ are the ghost and antighost
fields. The covariant derivative in the adjoint representation is
given by
\begin{equation}\label{eq4}
    D_\mu^{ab}=\delta^{ab}\p_\mu -gf^{abc}A_\mu^c\;.
\end{equation}
For $d=3$, the coupling $g^2$ becomes a dimensional quantity itself.
As a consequence, the three-dimensional Yang-Mills theory
(\ref{eq1}) is superrenormalizable. Such theories exhibit serious
problems in the infrared sector, as extensively discussed in
\cite{Jackiw:1980kv}. The origin of this problem is easy to
understand. If we were to write down a perturbative series in the
massive coupling $g^2$, then at a certain point inverse powers of
the external momenta will inevitably appear in order to obtain a
certain desired dimensionality. Consequently, at increasing order of
perturbation theory, the low momentum region becomes more and more
problematic to consider.

A natural solution to this problem might be the dynamical generation
of a mass $m$. If this were the case, a perturbative expansion in
the dimensionless parameter $\frac{g^2}{m}$ might emerge, ensuring a
safe infrared limit. A few mechanisms and techniques behind such a
mass generation are presented in for example
\cite{Jackiw:1995nf,Jackiw:1997jg,Eberlein:1998yk,Cornwall:1997dc,Karabali:1995ps,Karabali:1996je}

A common feature of the approaches of for example
\cite{Jackiw:1995nf,Jackiw:1997jg,Eberlein:1998yk,Cornwall:1997dc}
is that the necessary calculations are performed with a massive
version of Yang-Mills in a certain gauge. Essentially, a
nonperturbative mass is derived by constructing a gap equation, in a
particular approximation scheme, with nontrivial solutions. This
necessitates in fact a proof of the renormalizability of the massive
Yang-Mills theories in the specific gauges used \footnote{The exact
form of the invoked mass term is also not fixed. Several
possibilities exist, see \cite{Jackiw:1995nf,Jackiw:1997jg}.}, as
the employed gap equation should be at least renormalizable.

In \cite{Dudal:2004ch}, we have already shown that the Yang-Mills
action (\ref{eq1}) in the Landau gauge, supplemented with the parity
preserving mass term $\frac{1}{2}m^2 A_\mu^2$
\begin{equation}\label{eq5}
    S=S_\mathrm{YM}+S_\mathrm{GF+FP}+\int d^d x \frac{1}{2}m^2
    A_\mu^a A_\mu^a\;,
\end{equation}
is renormalizable to all orders using the algebraic renormalization
formalism \cite{Piguet:1995er}. This general proof was possible
since the action (\ref{eq5}) enjoys a certain number of Ward
identities, even in the massive case, allowing one to sufficiently
restrict the most general allowed counterterm, so that it can be
reabsorbed by a multiplicative redefinition of the fields and
parameters present in the original action \cite{Dudal:2004ch}. When
we define the bare fields and parameters by
\begin{eqnarray}
A_{0\mu }^{a} &=&Z_{A}^{1/2}A_{\mu }^{a}\,\,,\,\;\;\;c_{0}^{a}
=Z_{c}^{1/2}c^{a}\,,\,\,\,\;\;\;\;\;\;\; \nonumber \\
\;g_{0}~
&=&~Z_{g}g\,\,,\;\;\;\;\;\;\;\;\;\;\;\;m_{0}^{2}=Z_{m^{2}}m^{2}\;,
\nonumber \\
\overline{c}_{0}^{a} &=&Z_{\overline{c}}^{1/2}\overline{c}%
^{a}\,,\,\,\,\;\;\;\;\;\;\;\;\;b_{0}^{a}=Z_{b}^{1/2}b^{a}\;,
\label{eq6}
\end{eqnarray}
then it turns out that only $Z_A$ and $Z_g$ are independent, while
all the other renormalization factors can be expressed in terms of
these by the following relations,
\begin{eqnarray}
Z_{\overline{c}}\,&=&Z_{c}=Z_{g}^{-1}Z_{A}^{-1/2}\;,
\label{eq7}\nonumber\\
Z_{b}&=&Z_{A}^{-1}\;,\nonumber\\
Z_{m^{2}}~&=&~Z_{g}\,Z_{A}^{-1/2}\;,
\end{eqnarray}
which follow from the Ward identities \cite{Dudal:2004ch}. We recall
that the operator $\int d^3x A_\mu^2$ is BRST-invariant on-shell
\cite{Kondo:2001nq}.

This situation is completely similar to four dimensions
\cite{Dudal:2002pq}. We especially want to focus attention on the
fact that the mass $m$ does not renormalize independently.

Although the algebraic formalism enables us to prove the
renormalizability to all orders, it does not allow us to say
anything about the potential ultraviolet finiteness of the theory
(\ref{eq5}). In \cite{Dudal:2004ch}, we already pointed out that the
theory is ultraviolet finite at the one loop level, when dimensional
regularization is employed. In this paper, we extend this analysis
to two loop order, by explicitly verifying the absence of
ultraviolet divergences. Here, we want to stress that
\emph{individual} diagrams can be divergent. Nevertheless, when
taken together to obtain the renormalization factors, it turns out
that no divergent terms remain, and hence the theory is also
completely finite at two loop order. Notice that it would be
sufficient to consider the renormalization of for example the gluon
and ghost two point function, as only two independent
renormalization factors are needed.

By a power counting argument, we shall consequently deduce that the
theory is ultraviolet finite to any order.

To our knowledge, a complete proof of the ultraviolet finiteness of
three-dimensional Yang-Mills theories using a renormalizable gauge
and including a regulating mass compatible with a Slavnov-Taylor
identity \cite{Dudal:2004ch}, is still lacking so far.

Most works remained at the one loop level, not encountering any
divergences in three dimensions. As such, the question of
renormalizability is not really an issue. In \cite{Eberlein:1998yk},
the method of \cite{Jackiw:1995nf,Jackiw:1997jg} was extended to two
loop order. However, a kind of unitary gauge was employed, which
multiplicative renormalizability is not guaranteed.

Some gauges used in the literature are nonlocal
\cite{Jackiw:1995nf,Jackiw:1997jg}. Evidently, it would also remain
to be seen whether the nonappearance of divergences at one loop
remains the case at higher orders due to the nonlocality of the
gauge, preventing a straightforward analysis of the
renormalizability.

\section{Ultraviolet finiteness at one loop order.}
Let us first briefly review the ultraviolet finiteness at one loop
order. Let us begin with the one loop ghost-antighost self-energy.
As already pointed out in \cite{Dudal:2004ch}, using the the
transversality of the gluon propagator in the Landau gauge, it can
be checked that the one loop ghost self-energy is free from
ultraviolet divergences. As a consequence we have that
\begin{equation}
Z_{c}=Z_{\overline{c}}=1+\mathcal{O}\left(\frac{g^4}{m^2}\right)\;.
\label{1gh}
\end{equation}
Analogously, the one loop correction to the ghost-antighost-gluon
vertex is also finite. The same feature holds for the one loop
Feynman diagrams contributing to the four-gluon vertex, from which
it follows that
\begin{equation}
Z_{g}^{2}Z_{A}^{2}=1+\mathcal{O}\left(\frac{g^4}{m^2}\right)\;.
\label{ga}
\end{equation}
Moreover, from the equations (\ref{eq7}), we have
\begin{equation}
Z_{A}=1+\mathcal{O}\left(\frac{g^4}{m^2}\right)\;,  \label{1a}
\end{equation}
so that
\begin{equation}
Z_{g}=1+\mathcal{O}\left(\frac{g^4}{m^2}\right)\;.  \label{1g}
\end{equation}
We conclude that the theory is completely free from ultraviolet
divergences at one loop.
\section{Ultraviolet finiteness at two loop order.}
We comment briefly on the two loop computation of the
renormalization constants using the massive action (\ref{eq5}). It
is based on the same approach for renormalizing the Curci-Ferrari
model in four dimensions, \cite{Gracey:2001ma,Browne:2002wd}. There
QCD is fixed in the Curci-Ferrari gauge and the gluon and ghosts are
given a mass via an on-shell BRST invariant dimension two operator.
Therefore, this corresponds to a natural infrared regulator in a
renormalizable theory. For our three dimensional case the mass plays
a similar role as an infrared regulator. Hence we can apply the same
vacuum bubble expansion method as \cite{Gracey:2001ma,Browne:2002wd}
inspired by the earlier approach of
\cite{Misiak:1994zw,Chetyrkin:1997fm}. One can systematically
replace propagators involving an external momentum, $p$, with a
propagator involving only internal momenta by the exact identity
\begin{equation} \frac{1}{[(k-p)^2+m^2]} ~=~ \frac{1}{(k^2+m^2)} ~+~
\frac{(2kp - p^2)} {[(k-p)^2+m^2](k^2+m^2)} \;.
\end{equation}
This can be repeated iteratively with the truncation criterion being
determined by the particular Green function under consideration and
exploiting the renormalizability of the Lagrangian. So, for
instance, for  a gluon $2$-point function one expands to $O(p^2)$
since terms of higher order in the external momenta will be
ultraviolet finite by renormalizability and hence can be ignored for
the purposes of determining the renormalization constants. The
integrals which are retained involve vacuum bubbles which are
reduced to Lorentz scalar integrals after an elementary reduction of
the tensor structure in the numerator of the integrand. All that
remains is the evaluation of these three dimensional integrals at
two loops. This is achieved by the results summarized in
\cite{Rajantie:1996cw} where three dimensional massive vacuum
bubbles have been evaluated to {\em three} loops. More specifically
we note that all the (non-trivial) vacuum bubbles can be related to
the integral
\begin{equation} \int
\frac{d^dk d^d\ell}{(k^2+m_1^2)(\ell^2+m_2^2)[(k-\ell)^2+m_3^2]} ~=~
\frac{1}{(4\pi)^2} \left[ \frac{1}{4\varepsilon} ~+~ \frac{1}{2} ~+~
\ln \left[ \frac{\mu}{(m_1 + m_2 + m_3)} \right] \right] ~+~
O(\varepsilon)\;, \label{vacint2} \end{equation} where
$d$~$=$~$3$~$-$~$2\varepsilon$ in dimensional regularization and
$\mu$ is the mass scale introduced to maintain the same
dimensionality of the coupling constant in $d$ as in three
dimensions.

We have implemented the above algorithm in the symbolic manipulation
language {\sc Form}, \cite{Vermaseren:2000nd}, and automatically
renormalized all the $n$-point functions relevant for determining
all the renormalization constants of the Lagrangian. The Feynman
diagrams were generated automatically by the {\sc Qgraf} package,
\cite{Nogueira:1991ex}, and converted to {\sc Form} input notation.
The upshot of the renormalization is that all the renormalization
constants are unity up to and including two loops (in the $\MSbar$
scheme). At one loop this is virtually a trivial observation due to
the fact that there the basic master vacuum bubble is finite
\begin{equation} \int \frac{d^dk}{[k^2+m^2]} ~=~ -~ \frac{m}{4\pi}
\left[ 1 ~+~ \left[ 2 + 2 \ln \left( \frac{\mu}{2m} \right) \right]
\varepsilon \right] ~+~ O(\varepsilon^2) \;.
\end{equation}
This is yet another verification of the fact that the one loop
quantum corrections are finite.

At two loops the situation is less trivial since the basic integral
(\ref{vacint2}) is ultraviolet divergent but when all diagrams to
the Green functions involving this integral are added together then
the simple pole in $\varepsilon$ has zero residue. For instance, in
the renormalization of the ghost two point function $6$ of the $13$
two loop Feynman diagrams are divergent but overall the resulting
renormalization constant is unity.

\section{Ultraviolet finiteness at all orders.}
The general counterterm structure corresponding to the action
(\ref{eq5}) will look like
\begin{equation}\label{c1}
    S_\mathrm{c}=S_1 g^2+ S_2 g^4 + S_3 g^6+
    S_4\frac{g^8}{m}+\ldots\;,
\end{equation}
where the $S_i$ contains the counterterms proportional to powers of
$\frac{1}{\varepsilon}$. However, it is known that counterterms have
to be polynomial in the fields, their derivatives and the external
parameters like masses \cite{Collins:1974bg}, when massless
renormalization schemes like $\MSbar$ are employed. Therefore, all
coefficients $S_i$, $i>3$, should be zero. We can also forget about
the three loop term $\propto g^6$, as this is already of dimension
three, so it is merely a potentially infinite constant, but not
affecting any Green function. We come to the conclusion that the
only counterterms that can occur, must emerge at one or two loop
order. Having explicitly checked that there are no such
counterterms, we are able to conclude that the theory (\ref{eq5}) is
completely free of any ultraviolet divergence. As a consequence, all
Green functions can be unambigously calculated, without the need of
any renormalization scheme.

\section{The Curci-Ferrari gauge.}
The mass operator $A_{\mu }^{a}A_{\mu }^{a}$ in the Landau gauge can
be generalized to the Curci-Ferrari gauge, in which case the mixed
gluon-ghost mass operator $\left( \frac{1}{2}A_{\mu }^{a}A_{\mu
}^{a}+\alpha \overline{c}^{a}c^{a}\right) $ has to be considered,
where $\alpha $ stands for the gauge parameter. This operator also
enjoys the property of being BRST invariant on-shell. The action
corresponding to this nonlinear gauge fixing reads
\begin{eqnarray}
S &=&\int d^{3}x\,\left( -~\frac{1}{4}F_{\mu \nu }^{a}F_{\mu
\nu }^{a}+b^{a}\partial _{\mu }A_{\mu }^{a}+\,\frac{\alpha }{2}b^{a}b^{a}+%
\overline{c}^{a}\partial _{\mu }D_{\mu }^{ab}c^b-\frac{\alpha }{2%
}gf^{abc}b^{a}\overline{c}^{b}c^{c}\right.   \nonumber \\
&&\;\;\;\;\;\;\left. -\frac{\alpha }{8}g^{2}f^{abc}f^{cde}\overline{c}^{a}%
\overline{c}^{b}c^{d}c^{e}+m^{2}\left( \frac{1}{2}A_{\mu }^{a}A_{\mu
}^{a}+\alpha \overline{c}^{a}c^{a}\right) \right) \;.  \label{cf}
\end{eqnarray}
The Landau gauge follows from the Curci-Ferrari gauge in the limit
$\al\to0$.

It was proven in \cite{Dudal:2004ch} that (\ref{cf}) is
multiplicatively renormalizable to all orders. Due to the Ward
identities, it also turned out that the renormalization factor of
the mass $m$ is not an independent quantity of the theory either
\cite{Dudal:2004ch}. Moreover, we have checked that the results of
sections III and IV do also extend to the Curci-Ferrari case. In
other words, the action (\ref{cf}) describes an ultraviolet finite
theory.

\section{Conclusion.}
Having proven the renormalizability and in addition the ultraviolet
finiteness of the actions (\ref{eq5}) and (\ref{cf}), the question
is what happens in other gauges, which have also been supplemented
with a regulating mass that is consistent with renormalizability. We
expect that all the existing results from four-dimensional
Yang-Mills gauge theories should translate to three dimensions. We
refer to \cite{Dudal:2006xd,Lemes:2006aw} for the available
literature on several covariant gauges and their corresponding
renormalizable mass operators. Let us also mention that these gauges
and operators can be connected to each other by means of
renormalizable interpolating gauges. It remains an open problem if
the three-dimensional version of these mass regularized theories
would also be ultraviolet finite at all orders.

We have considered regulated three-dimensional gauge theories. Of
course, the real challenge would be to find a mechanism behind the
dynamical generation of a mass. In principle, as it concerns a gauge
theory, this mass should be obtained in a gauge invariant way. The
used mass operator, $\int d^3x A_\mu^2$ is apparently not gauge
invariant. Even more, it is not renormalizable in any gauge.
Nevertheless, it has a special feature in the Landau gauge worth
mentioning. Completely analogous as in four dimensions, $A_\mu^2$
can be related to the operator
\begin{equation}\label{con1}
    A^2_{\min} =(VT)^{-1}\min_{U\in SU(N)}\int d^3x \left(A_\mu^U\right)^2\;,
\end{equation}
which is gauge invariant due to the minimization along the gauge
orbit. $A^2_{\min}$ reduces to $A_\mu^2$ in the Landau gauge
\footnote{Finding the global minimum is however a delicate problem,
due to the existence of gauge (Gribov) copies.}. We refer to
\cite{Dudal:2006xd} for a list of references.

We also notice that the mass $m$ should be fixed. This should be
done by some kind of consistent gap equation, as $m$ is not supposed
to be a free parameter of the theory, see also
\cite{Jackiw:1995nf,Jackiw:1997jg,Sorella:2006ax}. This is however
beyond the scope of the present paper.

Nevertheless, it would be interesting to pursue the search for a
gauge invariant mass insertion. Inspired by
\cite{Jackiw:1995nf,Jackiw:1997jg}, we have started the
investigation of the gauge invariant mass operator $\int
d^4xF_{\mu\nu}^a\left[\left(D^2\right)^{-1}\right]^{ab}F_{\mu\nu}^b$
in four dimensions. In four dimensions, it is possible to add this
operator to the action as a mass term and to localize it in a way
consistent with the gauge invariance. Starting from this, a local,
non-Abelian gauge invariant action with a mass was constructed
\cite{Capri:2005dy,Capri:2006ne}. The renormalizability has been
proven to all orders in the linear covariant gauges. This study
could be extended to three dimensions, which might allow for a gauge
invariant discussion of a potential dynamical, parity preserving,
mass generation. An even more interesting question would be if this
action would also give rise to an ultraviolet finite theory.

\section*{Acknowledgments.}
The Conselho Nacional de Desenvolvimento Cient\'{i}fico e
Tecnol\'{o}gico (CNPq-Brazil), the Faperj, Funda{\c{c}}{\~{a}}o de
Amparo {\`{a}} Pesquisa do Estado do Rio de Janeiro, the SR2-UERJ
and the Coordena{\c{c}}{\~{a}}o de Aperfei{\c{c}}oamento de Pessoal
de N{\'\i}vel Superior (CAPES) are gratefully acknowledged for
financial support. D.~Dudal acknowledges the hospitality at the UERJ
where this work was completed.

\end{document}